\documentclass[journal]{IEEEtran}
\usepackage{amsmath}
\usepackage{graphicx}
\usepackage{xcolor}
\newcommand{\modified}[1]{\textcolor{black}{#1}}
\ifCLASSINFOpdf
\else
\fi
\usepackage{caption}
\usepackage{soul}
\usepackage{bm}
\usepackage{amssymb}
\usepackage{algorithm}
\usepackage{algpseudocode}
\usepackage{multirow}
\usepackage{xcolor}
\usepackage[inline]{enumitem}
\usepackage{mathtools}
\definecolor{steelblue}{RGB}{48 126 167}
\usepackage{varioref}
\usepackage{url}
\usepackage{hyperref}
\hypersetup{
  colorlinks = true,   % Führt zu einem farbigen Ausdruck!
  linkcolor =  black,
  urlcolor =   steelblue,
  citecolor =  black,
  plainpages =        false,
  hypertexnames =     true,
  linktocpage =       true,
  bookmarksopen =     true,
  bookmarksnumbered = true,
  bookmarksopenlevel= 0,
}
\usepackage[all]{hypcap}
\usepackage[noabbrev,capitalise,nameinlink]{cleveref}
\definecolor{lightorange}{RGB}{255, 204, 153}
\usepackage{cite}
\begin{document}
\title{Enhancing Cross-Modality Synthesis: Subvolume Merging for MRI-to-CT Conversion}

\author{Fuxin~Fan,
        Jingna~Qiu,
        Yixing~Huang,
        and~Andreas~Maier% <-this % stops a space ~\IEEEmembership{Member,~IEEE,}
\thanks{F.~Fan and A.~Maier are with the Pattern Recognition Lab, Friedrich-Alexander-Universität Erlangen-Nürnberg,
Germany. J.~Qiu is with the Department of Artificial Intelligence in Biomedical Engineering, Friedrich-Alexander-Universität Erlangen-Nürnberg,
Germany. Y.~Huang is with the Department of Radiation Oncology, University Hospital Erlangen, Friedrich-Alexander-Universität Erlangen-Nürnberg, Germany. Contact E-mail: (fuxin.fan@fau.de).}}% <-this % stops a space
\maketitle

\begin{abstract}
% With more accurate tissue attenuation information, the synthetic computed tomography (sCT) from the magnetic resonance imaging (MRI) allows for better radiation therapy treatment planning. In this work, a state-of-the-art framework, named Swin UNETR, is utilized for CT synthesis from MRI images. In addition, three-dimentional subvolume merging is proposed during prediction process. By choosing a proper overlap percentage between subvolumes, the checkerboard artifacts are effectively eliminated, and the mean absolute error (MAE) between sCT and labels can be reduced from 52.65 HU to 47.75 HU. Furthermore, a weight function with gamma value of 0.9 achieves the lowest MAE under the same overlap area. By choosing the overlap percentage between 50\% and 70\%, a performance balance between the image quality and computational time can be achieved.
Providing more precise tissue attenuation information, synthetic computed tomography (sCT) generated from magnetic resonance imaging (MRI) contributes to improved radiation therapy treatment planning. In our study, we employ the advanced SwinUNETR framework for synthesizing CT from MRI images. Additionally, we introduce a three-dimensional subvolume merging technique in the prediction process. By selecting an optimal overlap percentage for adjacent subvolumes, \modified{stitching} artifacts are effectively mitigated, leading to a decrease in the mean absolute error (MAE) between sCT and the labels from 52.65 HU to 47.75 HU. Furthermore, implementing a weight function with a gamma value of 0.9 results in the lowest MAE within the same overlap area. By setting the overlap percentage between 50\% and 70\%, we achieve a balance between image quality and computational efficiency.
\end{abstract}


% \fi
\IEEEpeerreviewmaketitle

\thispagestyle{empty}

\section{Introduction}
% \IEEEPARstart{C}{omputed} Tomography (CT) and magnetic resonance imaging (MRI) are widely used in the radiotherapy workflow. On the one hand, accurate dose calculations for the workflow rely on electron density maps generated from CT images. On the other hand, MRI images provide superior soft tissue contrast over CT, which contribute to the treatment planning by providing better target segmentation. Therefore, it is necessary to perform registration from MRI to CT. However, registration errors could exist, which will degrade the reliability of treatment. To avoid registration errors, CT synthesis from MRI is a possible solution. In addition, using synthetic CT (sCT) rather than acquiring CT scans will avoid extra patient radiation exposure and reduce the treatment cost at the same time.

\IEEEPARstart{C}{omputed} Tomography (CT) and Magnetic Resonance Imaging (MRI) play crucial roles in radiotherapy planning \cite{fu2020generation}. CT images are essential for generating accurate electron density maps, which are vital for precise dose calculations \cite{brou2021improving}. Conversely, MRI is effective at offering superior soft tissue contrast, enhancing target segmentation in treatment planning \cite{dayarathna2023deep}. Consequently, aligning MRI with CT images (MRI-to-CT registration) is a critical step. However, this process is susceptible to registration errors, potentially compromising treatment reliability \cite{lei2019mri}. To circumvent the MRI-to-CT registration errors, synthesizing CT from MRI images emerges as a viable alternative. Moreover, employing synthetic CT (sCT) instead of traditional CT scans can reduce additional radiation exposure to patients and lower treatment costs \cite{owrangi2018mri}.

% With the development of deep learning, many MRI-to-CT conversion algorithms have been proposed. The U-Net, which has a symmetrical architecture, is a popular and efficient network for CT synthesis. Various generative adversarial networks (GANs), which introduce discriminative loss, are also developed for MRI-to-Ct translation task. With the attention mechanism, vision transformer (ViT) based networks show their good performance in this task. Recently, a challenge named SynthRad2023 provides registered MRI and CT images from different institutions for CT synthesis task, which makes this topic accessible for researchers all over the world.

% However, previous work mainly focuses on network design but neglects the postprocessing stage, which could further improve the quality of the prediction. In this paper, we propose to use subvolume training for CT synthesis from MRI, employing a state-of-the-art structure, named Swin UNETR. More importantly, effective subvolume merging strategy is proposed in the prediction process.

With the development of deep learning technologies, a diverse range of MRI-to-CT conversion algorithms has been proposed. The U-Net, known for its symmetrical design, stands out as an effective model for CT synthesis \cite{spadea2019deep,dovletov2022grad}. Additionally, various Generative Adversarial Networks (GANs), which incorporate discriminative loss, have been developed specifically for MRI-to-CT translation \cite{lei2019mri,qian2020estimating,sun2023double}. Networks based on Vision Transformers (ViT) with attention mechanisms have also demonstrated impressive performance in this area \cite{li2023ct,zhao2023ct}. The recent SynthRad2023 challenge has contributed to this field by making registered MRI and CT images from multiple institutions available to researchers over the world \cite{thummerer2023synthrad2023,adrian_thummerer_2023_7781049}.

Despite these developments, most existing studies have primarily concentrated on network architecture, often overlooking the potential enhancements in the postprocessing stage. Our paper addresses this gap by introducing a novel approach to CT synthesis from MRI through subvolume training, utilizing the SwinUNETR structure \cite{hatamizadeh2021swin}. Crucially, we present an innovative subvolume merging strategy during the prediction phase, aiming to further enhance the quality of the synthesized CT images.

\section{Methods and Materials}

\subsection{Neural network}
\begin{figure}[htb]
    \begin{center}
    \begin{tabular}{c} %% tabular useful for creating an array of images 
    \includegraphics[width=8.5cm]{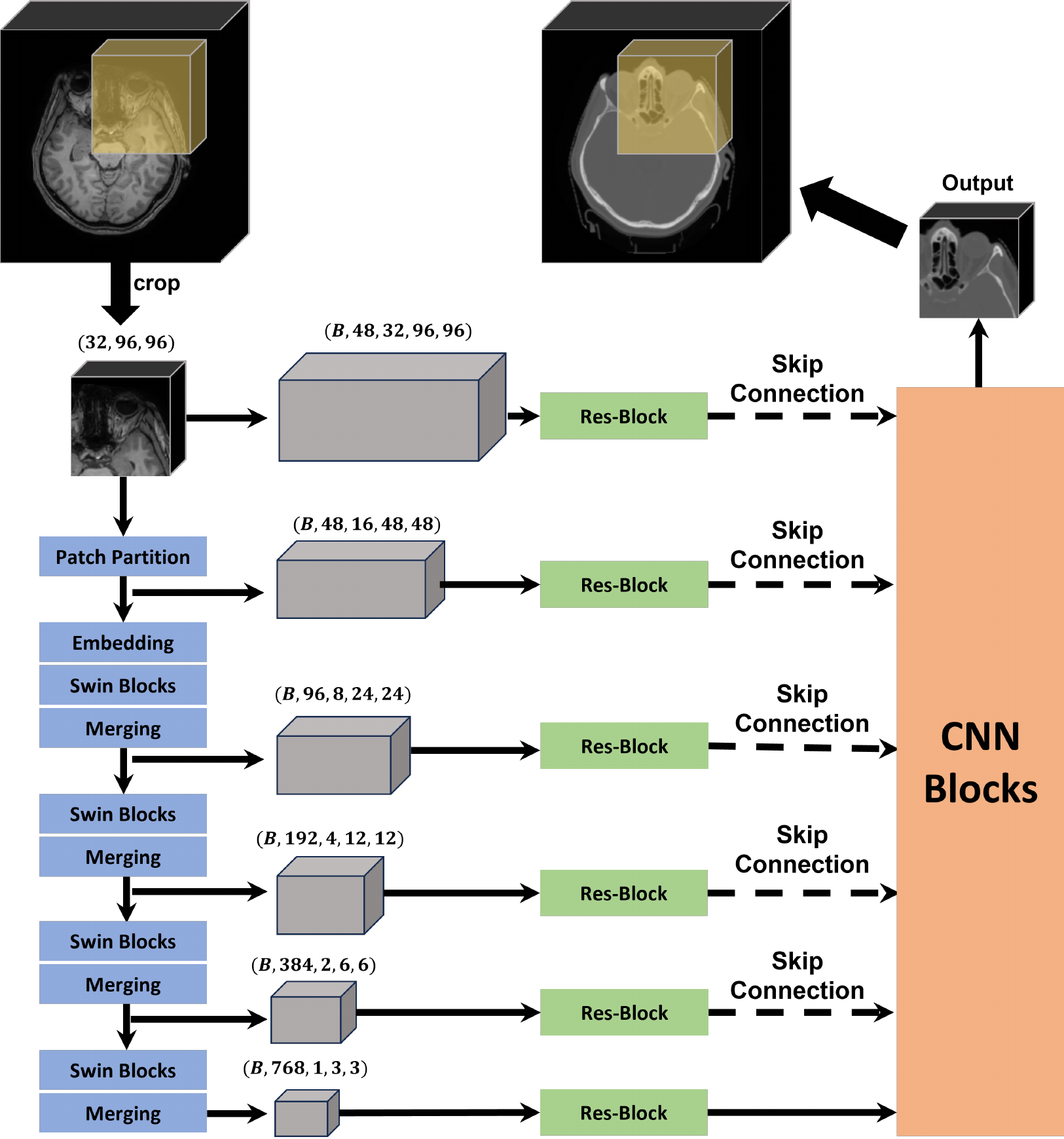}
    \end{tabular}
    \end{center}
    \captionsetup{justification=centering} 
    \caption
    {The network structure of SwinUNETR.}
    \label{fig:network}
\end{figure}

\begin{figure*}[htb!]
    \begin{center}
    \begin{tabular}{c} %% tabular useful for creating an array of images 
    \includegraphics[width=15cm]{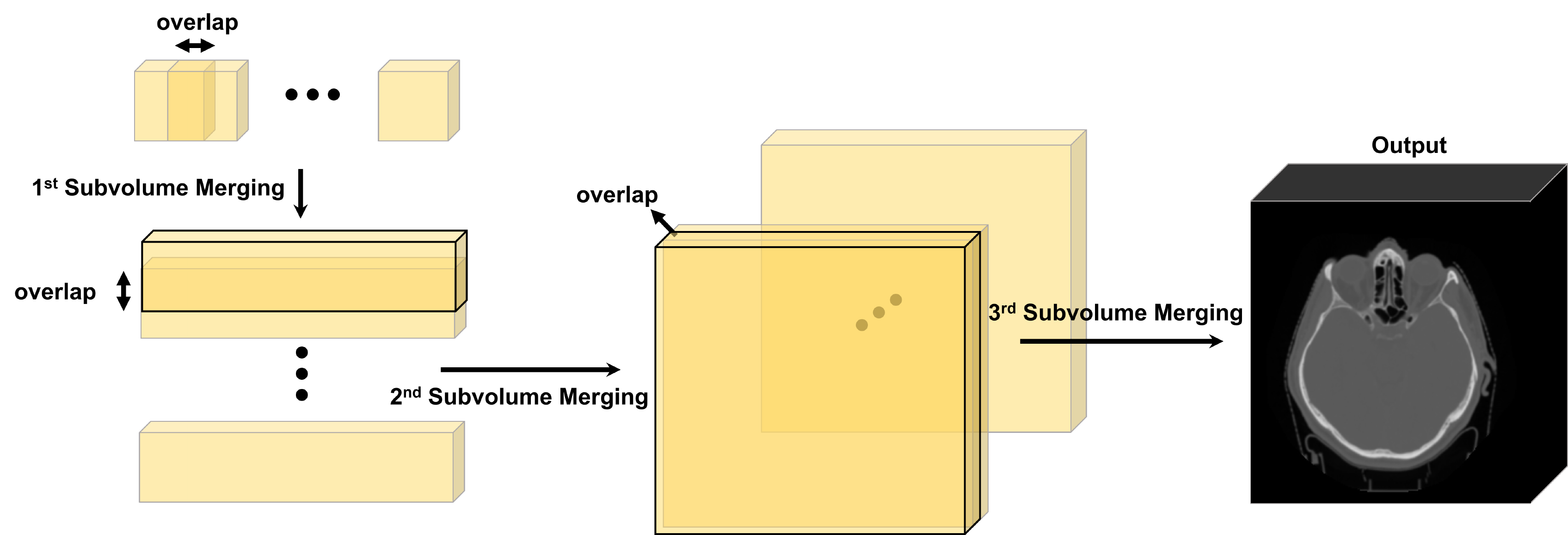}
    \end{tabular}
    \end{center}
    \captionsetup{justification=centering}    
    \caption
    {3D subvolume merging during the inference process.}
    \label{fig:merging}
\end{figure*}
In this work, we use a state-of-the-art network, SwinUNETR, for MRI-to-CT synthesis. The implementation of the SwinUNETR is available under the open-source framework MONAI \cite{cardoso2022monai}. The architecture of the SwinUNETR is shown in Fig.~\ref{fig:network}. This network consists of a Shift window (Swin) Vision transformer (ViT)-based encoder and a CNN-based decoder.
A subvolume of size $32\times96\,\times\,96\,$ is randomly selected from an MRI volume and fed into the network. The SwinUNETR splits the subvolume into a sequence of patches, with the size of $2\times2\,\times\,2\,$. Each patch is embedded into a vector with the feature dimension of 48. The patch sequence then goes through four stages, and each stage has 2 Swin blocks followed by a patch merging operation. After patch merging, the side length of one patch is doubled. At the same time, the output dimension is also doubled. The output from each stage is reshaped and forwarded into a residual block before concatenating with CNN-based blocks. The residual block consists of two $2\times2\,\times\,2\,$ convolutional layers followed by an instance normalization layer. In each CNN block, the concatenated features are fed into another residual block and a deconvolutional layer. The feature size gets halved after the deconvolutional layer. The final outputs with a single channel are computed by using a $1\times1\,\times\,1\,$ convolutional layer.

The network was trained using an NVIDIA A100 GPU with 80 GB memory. The model is trained on 144 patient cases in each epoch. For each case, 20 subvolumes are randomly selected for each epoch. The predictions and labels are pixel-wisely multiplied with their corresponding binary masks before loss calculation. The L1 loss function and the Adam optimizer are used. The values for $\beta_{1}$ and $\beta_{2}$ are 0.9 and 0.999. The models are trained for maximum 1000 epochs, and the training stops when the validation loss is not decreased for three continuous epochs. The learning rate has stepwise decay from 0.0005 to 0.00005.

\thispagestyle{empty}

\subsection{Subvolume merging}
To reduce the inference time, only subvolumes within binary masks are predicted. The whole CT volume is constructed by merging adjacent subvolumes sequentially in three dimensions. The merging process is shown in Fig.~\ref{fig:merging}. The smallest subvolumes construct long cuboids. Then long cuboids are connected with each other to build flat cubes. The CT volume is then obtained by merging all flat cubes together. The overlap areas of adjacent subvolumes are multiplied with two weight maps to keep smooth intensity transition. For the same voxel from two adjacent subvolumes a and b, the merged intensity $I$ is the weighted summation of $I_{A}$ and $I_{B}$:
\begin{equation}
\label{equ:weight}
    I_{j} = (1-w_{j})I_{j,A} + w_{j}I_{j,B}.
\end{equation}
The weight $w_{j}$ satisfies an exponential function:
\begin{equation}
    w_{j} = (j/N)^{\gamma},
\end{equation}
where $N$ is the overlap length along one specific direction \modified{and $j$ is the voxel index along the overlap direction. When $\gamma =1$, a linear weight is used.} As shown in Eq.~(\ref{equ:weight}), the weight for the former subvolume $A$ decreases from 1 to 0 along the direction, whereas the weight for the latter increases from 0 to 1 complementarily. In addition, the overlap percentage ranging between 0 and 1 is defined as the ratio between the overlap length and the image length. 

\subsection{Data set preparation}
% In this work, 180 pairs of MRI and CT brain volumes from three different institutes are split into 144 cases for model training, 18 cases for validation and 18 cases for testing. Each volume from brain and neck region has been preprocessed to ensure the same voxel size of $1\,mm\times1\,mm\times\,1\,mm$. In addition, MRI and CT pairs are registered by organizers. All MRI intensity values are divided by 1000. CT values are subtracted by the minimum value of each volume (e.g., mostly -1024) to be nonnegative, and then divided by 2000. Afterwards, MRI subvolumes with the size of $32\times96\,\times\,96\,$, together with their corresponding binary masks and CT subvolumes are randomly selected to construct the data for network training. In addition, MRI subvolumes are pixel-wisely multiplied with their binary masks as the input of the network. Binary masks for all cases are provided to give the patient outline segmentation, and the regions within the segmentation are used for evaluation. Mean absolute error (MAE) and the peak-signal-to-noise-ratio (PSNR) are used as evaluation metrics.

In this study, we utilized 180 paired MRI and CT brain volumes sourced from three different institutions \cite{thummerer2023synthrad2023,adrian_thummerer_2023_7781049}. These were divided into 144 cases for training the model, 18 for validation, and 18 for testing. To standardize the data, each volume from the brain and neck region was preprocessed to achieve a uniform voxel size of 1\,mm $\times$ 1\,mm $\times$ 1\,mm. Furthermore, MRI and CT pairs were pre-registered by the organizers.

For data normalization, MRI intensity values were scaled down by a factor of 1000. CT values were first adjusted to be nonnegative by subtracting the minimum value of each volume (usually around -1000 HU) and then scaled down by a factor of 2000.

Subsequently, MRI subvolumes of size $32\times96\,\times\,96\,$, along with their corresponding binary masks and CT subvolumes, were randomly selected for network training. The MRI subvolumes were pixel-wisely multiplied with their binary masks before being fed into the network. These binary masks, provided for all cases, outline the patient's segmentation and the areas within these segmentations were the focus of our evaluation. We employed Mean Absolute Error (MAE) and Peak Signal-to-Noise Ratio (PSNR) as our primary evaluation metrics.

\section{Results and discussion}
\begin{figure*}[htb]
    \begin{center}
    \begin{tabular}{c} %% tabular useful for creating an array of images 
    \includegraphics[width=\textwidth]{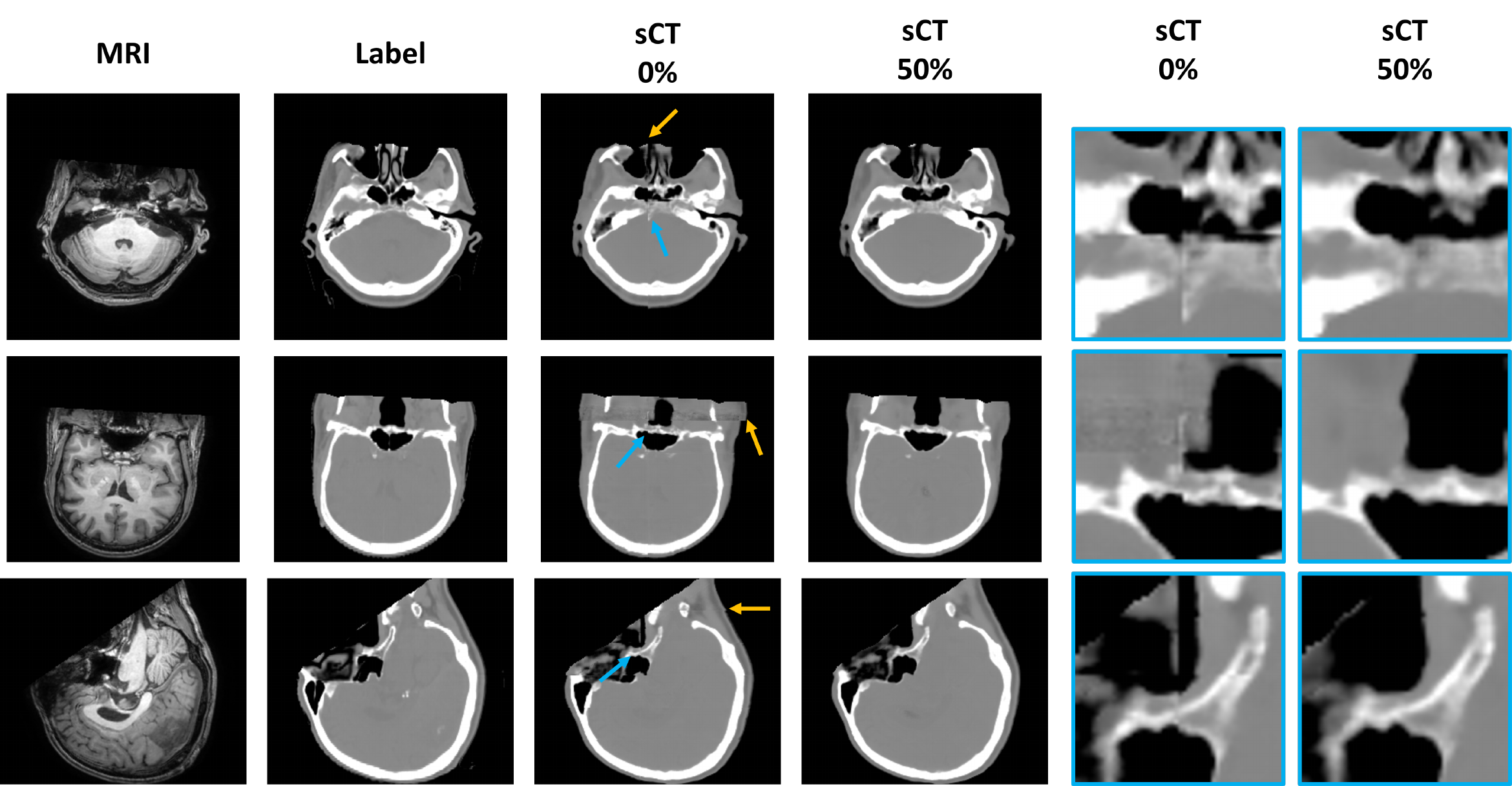}
    \end{tabular}
    \end{center}
    \captionsetup{justification=centering}    
    \caption
    {The result of sCT before and after subvolume merging. The right two columns show the enlarged regions from the 3rd and 4th columns. Such regions are highlighted by blue arrows in the third column. The intensity window for CT images is [-500, 500] HU.}
    \label{fig:result}
\end{figure*}
% The sCT images generated by subvolume merging with 0\% and 50\% overlap percentage are shown in Fig.~\ref{fig:result}. When the overlap percentage is 0\%, checkerboard artifacts highlighted by orange arrows can be observed in the predictions. On the contrary, such artifacts are eliminated when the overlap percentage is set to 50\%.
Fig.~\ref{fig:result} illustrates the synthetic CT (sCT) images produced through subvolume merging with 0\% and 50\% overlap percentages. Notably, \modified{stitching} artifacts, indicated by orange and blue arrows, are visible in the images with 0\% overlap. In the enlarged regions near blue arrows, such artifacts are more distinctive. In contrast, increasing the overlap percentage to 50\% effectively eliminates these artifacts.

\thispagestyle{empty}

% The curve in Fig.~\ref{fig:gamma} shows the change of MAE under different $\gamma$ value. This experiment was conducted when the overlap percentage was set to 50\%. With the increment of $\gamma$, the MAE decreases in the beginning and then increases afterward. The minimum MAE is achieved when $\gamma$ is set to 0.9.

\begin{figure}[htb]
    \begin{center}
    \begin{tabular}{c} %% tabular useful for creating an array of images 
    \includegraphics[width=8.5cm]{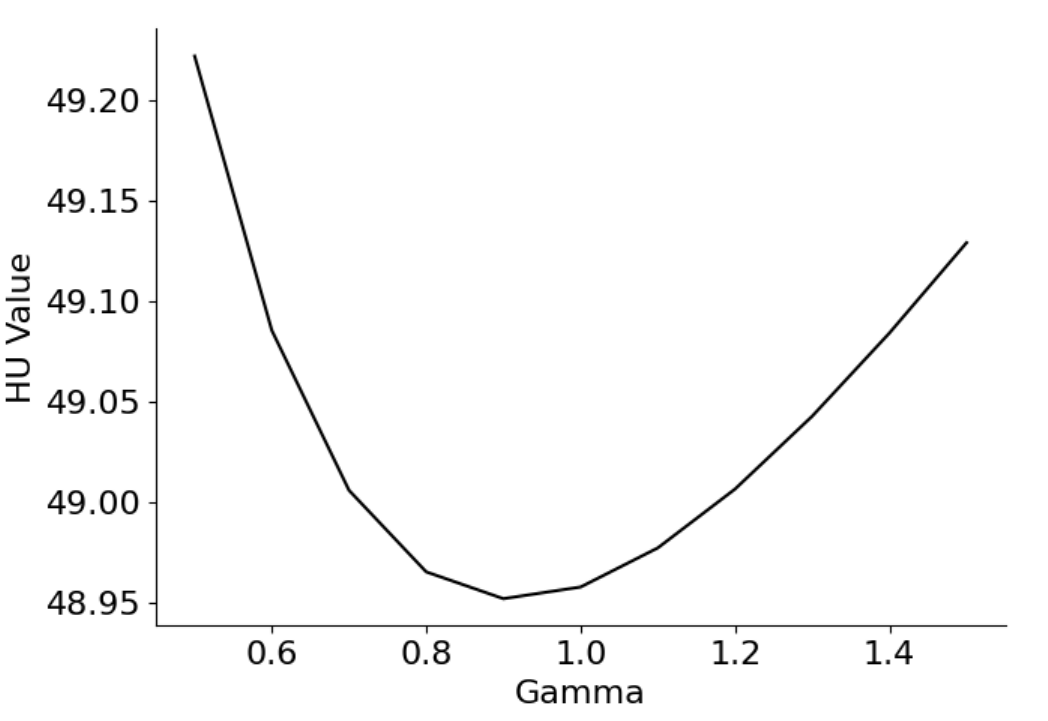}
    \end{tabular}
    \end{center}
    \captionsetup{justification=centering}    
    \caption
    {MAE with respect to Gamma values.}
    \label{fig:gamma}
\end{figure}

The graph displayed in Fig.~\ref{fig:gamma} depicts how the MAE varies with different values of $\gamma$. This analysis was performed with the overlap percentage fixed at 50\%. Initially, as $\gamma$ increases, there is a decrease in the MAE. However, beyond a certain point, the MAE begins to rise. The lowest MAE is observed when $\gamma$ is set to 0.9.

% Fig.~\ref{fig:overlap} illustrates the change of image metrics and the average number of subvolumes for testing cases with respect to the overlap percentage. In general, MAE decreases and PSRN increases with larger overlap percentage, which reflects an improvement on image quality. More specifically, the MAE is decreased from 52.6 HU to 47.75 HU, and the PSNR is increased from 27.84 to 28.65. However, the average number of subvolumes is enlarged for more than 600 times from 53 to 36318, which magnitudes the computational time during inference.

\begin{figure}[htb]
    \begin{center}
    \begin{tabular}{c} %% tabular useful for creating an array of images 
    \includegraphics[width=8.5cm]{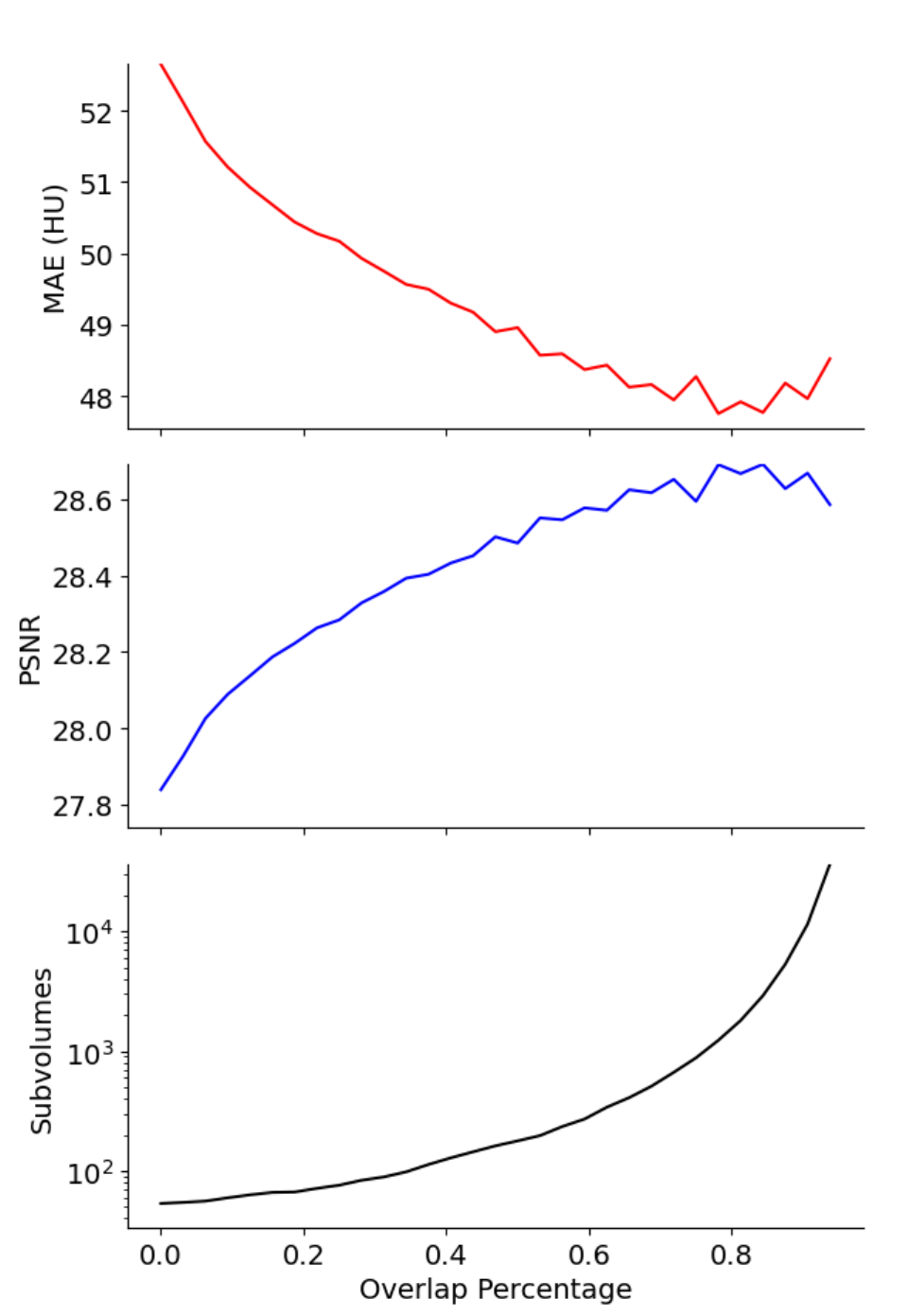}
    \end{tabular}
    \end{center}
    \captionsetup{justification=centering}    
    \caption
    {Ablation study on overlap percentage.}
    \label{fig:overlap}
\end{figure}

Fig.~\ref{fig:overlap} presents how image metrics and the average number of subvolumes for test cases vary with different overlap percentages. Generally, as the overlap percentage increases, there is a noticeable improvement in image quality: the MAE drops from 52.65 HU to 47.75 HU, and the PSNR rises from 27.84 to 28.65. However, this increase in overlap percentages also leads to a considerable rise in the average number of subvolumes, expanding from 53 to 36,\,318, thereby considerably increasing the computational time required for inference.

% \begin{figure}[htb]
%     \begin{center}
%     \begin{tabular}{c} %% tabular useful for creating an array of images 
%     \includegraphics[width=8.5cm]{figures/Result.pdf}
%     \end{tabular}
%     \end{center}
%     \captionsetup{justification=centering}    
%     \caption
%     {The result of sCT before and after subvolume merging. The intensity window for CT images is [-500, 500].}
%     \label{fig:result}
% \end{figure}

% When the models are trained on subvolumes, subvolume merging makes a difference for the final inference result. Such operation improves the image quality both quantitatively and qualitatively, which will benefit the following radiotherapy treatment planning. However, a high overlap percentage results in booming computational time. Therefore, it is necessary to reach a balance between image quality and inference time. According to our experiments, setting the overlap percentage between 50\% and 70\% is an ideal choice. Within this overlap percentage range, the MAE is between 48.12 HU and 48.96 HU, which has very little difference to each other. In addition, the number of subvolumes is between 178 and 508, which are 3.5 times and 9.5 times larger than the non-overlapping situation.

When models are trained using subvolumes, the process of subvolume merging plays a crucial role in shaping the final inference results. This method enhances image quality in both measurable and observable terms, which is advantageous for subsequent radiotherapy treatment planning. However, choosing a high overlap percentage can lead to an increase in computational time. Thus, finding a balance between image quality and inference speed is essential. Our experiments indicate that an overlap percentage between 50\% and 70\% is optimal. In this range, the MAE shows minimal variation, ranging from 48.12 HU to 48.96 HU. Moreover, the number of subvolumes falls between 178 and 508, which is 3.5 to 9.5 times bigger than that without overlap, but still manageable in terms of computational demand.

\section{Conclusion}
% In this work, we propose to use subvolume merging during the inference process, which generates superior images over inference by the naive subvolume joining. With the choice of 50\% and 70\% overlap percentage, the computational time is constrained while preserving the high image quality. Our strategy can be further applied in other regression applications which needs subimage or subvolume training.
In this study, we introduce the use of subvolume merging in the inference process, a technique that produces higher quality images compared to subvolume joining methods. By opting for an overlap percentage between 50\% and 70\%, we are able to maintain computational efficiency while ensuring high image quality. This approach has broader applicability and could be effectively implemented in other regression tasks that require subimage or subvolume training.

\thispagestyle{empty}

\section*{Acknowledgment}

The authors acknowledge the scientific support and HPC resources provided by the Erlangen National High Performance Computing Center (NHR@FAU) of the Friedrich-Alexander-Universität Erlangen-Nürnberg (FAU).

\ifCLASSOPTIONcaptionsoff
  \newpage
\fi

\bibliographystyle{IEEEtran}
\bibliography{ref}

\end{document}